\documentclass[12pt]{article}

\newcommand{\be}{\begin{equation}}
\newcommand{\ee}{\end{equation}}

\begin{document}

\newpage
\bigskip
\hskip 3.7in\vbox{\baselineskip12pt
\hbox{BROWN-HET-1167}
\hbox{hep-th/9902088}}

\bigskip\bigskip
\begin{center}
{\large \bf Holography, Cosmology \\and the\\
 Second Law of Thermodynamics}
\end{center}

\bigskip\bigskip

\centerline{\bf
Richard Easther\footnote{easther@het.brown.edu} and 
David A. Lowe\footnote{lowe@het.brown.edu}} 
\medskip
\centerline{Department of Physics}
\centerline{Brown University}
\centerline{Providence, RI 02912, USA}

\bigskip

\begin{abstract}
\baselineskip=16pt
We propose that in time dependent backgrounds the holographic
principle should be replaced by the generalized second law of
thermodynamics.  For isotropic open and flat universes with a fixed
equation of state, the generalized second law agrees with the
cosmological holographic principle proposed by Fischler and
Susskind. However, in more complicated spacetimes the two proposals
disagree. A modified form of the holographic bound that applies to a
post-inflationary universe follows from the generalized second
law.  However, in a spatially closed universe, or inside a black hole
event horizon, there is no simple relationship that connects the area
of a region to the maximum entropy it can contain.

\end{abstract}
\newpage
\baselineskip=18pt
\setcounter{footnote}{0}

\section{Introduction}

The holographic principle proposes that the maximum number of degrees of
freedom in a volume is proportional to the surface area
\cite{thooft,Susskind1995a}. This principle is based on earlier
studies by Bekenstein \cite{Bekenstein} of maximum entropy bounds
within a given volume.  One argument used to motivate the holographic
principle is as follows.  Consider a region of space with volume $V$,
bounded by an area $A$, which contains an entropy, $S$, and assume
that this entropy is larger than that of a black hole with the same
surface area.  Now throw additional energy into this region to form a
black hole.  Assuming that the Bekenstein-Hawking formula, $S =A/4$,
actually gives the entropy of the black hole, we conclude that the
generalized second law of thermodynamics \cite{Bekenstein1973a} has
been violated.  To avoid this contradiction, the holographic principle
proposes that the entropy inside a given region must satisfy $S/A <
1$.  However, this line of reasoning implicitly assumes that the black
hole forms in an otherwise static background.

In what follows, we examine how this argument changes in more general,
time-dependent, spacetimes, such as those encountered in cosmology.
We argue the holographic bound is replaced by the simple requirement
that physics respects the generalized second law of thermodynamics
\cite{Bekenstein1973a}.  For static backgrounds this reduces to the
holographic bound but, in general, the maximum entropy permitted
inside a region is not related to its area by a simple formula.

Fischler and Susskind \cite{FischlerET1998a} have proposed a
generalization of the holographic principle to certain cosmological
backgrounds.   This proposal has been studied further in
\cite{otherhol, RamaET1998a}.  For flat and open universes with time
independent equations of state, we find that their bound is in accord
with the generalized second law, and we propose a refinement of the
Fischler-Susskind bound that applies to an inflationary universe after
reheating.

Fischler and Susskind found that closed universes violate their
cosmological holographic bound, and speculated that such backgrounds
were either inconsistent, or that new behavior sets in as the bound is
violated. We argue that the evolution of closed universes does not
violate the generalized second law, and that such backgrounds are thus
self-consistent.

A related problem is the application of the holographic principle to a
volume inside the event horizon of a black hole. The na\"{\i}ve
holographic bound can easily be violated in such a region. Although
this evolution respects the generalized second law, it appears that
the price an observer pays for violating the holographic bound is to
eventually encounter a curvature singularity. However it is possible
for this fate to be delayed for cosmologically long time scales.

\section{The Story So Far}

Fischler and Susskind \cite{FischlerET1998a} realized that while the
holographic bound, $S/A < 1$, applies to an arbitrary region for the
static case, its application to cosmological spacetimes is more
subtle. Specifically, the homogeneous energy density, $\rho$, of
simple cosmological models implies a homogeneous entropy density,
$s$. Inside a (spatial) volume $V \sim R^3$, the total entropy is $S =
sV$, but $S/A \sim s R$.  Consequently, for a fixed $s$ it is always
possible to choose a volume large enough to violate the holographic
bound.  Fischler and Susskind resolve this problem by stipulating that
the holographic bound only applies to regions smaller than the
cosmological (particle) horizon \cite{Rindler1956a}, which corresponds
to the forward light-cone of an event occurring at (or infinitesimally
after) the initial singularity.  The comoving distance to the horizon,
$r_H$, is
\be 
r_h = \int_0^{t_0}{\frac{1}{a(t')}dt'}
\ee
while the corresponding physical distance is
\be 
d_h = a(t) r_h = a(t) \int_0^{t_0}{\frac{1}{a(t')}dt'}.
\ee
Here $a(t)$ is the scale factor of the Robertson-Walker metric, and
obeys the evolution equations
\begin{eqnarray}
\left(\frac{\dot{a}}{a}\right)^2 = H^2 &=& \frac{\rho}{3 } -
\frac{k}{a^2} \label{rw1}
\\
\frac{\ddot{a}}{a} &=& -\frac{ \left(\rho + 3 p \right)}{6}
\label{rw2}
\end{eqnarray}
where $k$ takes the values $\pm1$ and 0, for solutions with positive,
negative and zero spatial curvature.

For a perfect fluid, in a flat ($k=0$) universe, whose pressure and
density satisfy $\rho = \omega p$, the solution of equations
(\ref{rw1}) and (\ref{rw2}) is straightforward
\be
a(t) = a_0 \left(\frac{t}{t_0}\right)^q, \qquad
q = \frac{2}{3} \frac{1}{1+\omega}.
\ee
In particular, if $\omega=0$ we recover the equation of state for
dust, while $\omega=1/3$ is the appropriate value for a hot
(relativistic) gas or radiation.  In general,
\be
d_H = \frac{t}{1-q}~.
\ee
The comoving entropy density is constant, so with $k=0$ it follows
that when measured over the horizon volume,
\be
\frac{S}{A} \propto t^{1-3q}.
\ee
If $q< 1/3$ ($\omega > 1$) the holographic bound is violated at late
times but, as Fischler and Susskind explain, such a cosmological model
is not viable since a perfect fluid with $\omega>1$ has a speed of sound
greater than the speed of light.

In realistic cosmological models the equation of state is far from
that of a perfect fluid with constant $\omega$.  Even simple models of
the big bang combine dust and radiation and make a transition between
$\omega = 1/3$ and $\omega = 0$, since the energy density of radiation
drops faster than the density of dust as the universe expands. More
importantly, during an inflationary epoch in the primordial universe
$\ddot{a}$ is, by definition, positive; so the pressure and $\omega$
must be negative.

One of the original motivations for inflation was that it endows the
primordial universe with a substantial entropy density.  Inflationary
models generate entropy after inflation has finished, when energy is
transferred from the scalar field which drove the inflationary
expansion to radiation and ultra-relativistic particles.  This process
is referred to as reheating, and the equation of state usually changes
from $\omega < -1/3$, to that of a radiation dominated universe whose
subsequent evolution is described by the ``standard'' model of the hot
big bang.  The comoving entropy density is not constant, and $S/A$ is
thus a more complicated function of time than it is in models with
constant $\omega$.

The maximum temperature, $T$, attained after inflation is model
dependent, and the resulting entropy density is proportional to $T^3$
only if we assume a relativistic gas.  Inflation can make the
cosmological horizon arbitrarily large; for instance in a class of
realistic models it may be $10^{1000}$ times greater than in the
absence of inflation.  Applying the original Fischler-Susskind
formulation of the holographic principle leads to a value of $S/A$
massively greater than unity for almost any realistic inflationary
model.  This difficulty is noted by Rama and Sarkar
\cite{RamaET1998a}, and they propose various smaller volumes over
which to measure the entropy. In general, their formulation is not
consistent with the one we propose in the next section.

\section{Holography and the \\ Generalized Second Law}

One of the initial motivations for the holographic principle was based
on the generalized second law of thermodynamics. The generalized
second law states
\be
\delta S_{mat} + \delta S_{bh} \ge 0~,
\ee
where $S_{mat}$ is the entropy of matter outside black holes, and
$S_{bh}$ is the Bekenstein-Hawking entropy of the black holes. This
law has not been proven, but is expected to follow from most of the
current approaches to quantum gravity and has been tested in many
non-trivial situations \cite{Bekenstein}.  Assuming that this law is
correct, the holographic principle follows from it if we consider a
region of space embedded in an approximately static background (such
as Minkowski space, or anti-de Sitter space), as discussed in the
introduction.

Our main interest is to study the formulation of holographic style
bounds in a general background.  In time dependent situations, we
propose that the general principle which replaces the holographic
principle is simply: {\it that the generalized second law of
thermodynamics holds.}

In the examples considered below, we assume that changes are
quasi-static, so that at all times the entropy is maximized to an
arbitrarily good approximation, subject to constraints.  In these
situations we can make the stronger statement: {\it for all time, the
entropy is maximized subject to the constraints.}

For volumes embedded in certain backgrounds we may use these
principles to deduce holographic style bounds on the entropy, but 
this does not appear to be possible in general.  We will now
illustrate these observations with a number of examples.

\subsection{Flat Universe}

Let us consider isotropic, homogeneous and spatially flat cosmologies.
The comoving entropy density in these models is
constant.%
\footnote{This discussion, like that of Fischler and Susskind, assumes
that a flat or open universe necessarily expands indefinitely, while a
closed universe recollapses in a finite time.  However, a spatially
flat or open universe with a negative vacuum energy density
(cosmological constant) can recollapse, just as a positive vacuum
energy can cause a closed universe to expand indefinitely. Our
discussion can easily be generalized to these cases.}
In order to formulate a holographic bound, we need to introduce a
length scale that defines the size of the spatial region under
consideration. We argue that the relevant length scale is the Hubble
length $1/H$.  Physically, $1/H$ is the distance at which a point
appears to recede at the speed of light, due to the overall expansion
of the universe.

To see that this is the relevant length scale consider a small
gravitational perturbation of this background. Small perturbations to
a spatially flat, homogeneous and isotropic universe with wavelengths
larger than $1/H$ do not grow with time \cite{MukhanovET1992b},
provided the equation of state remains constant.  If perturbations do
not grow, black holes cannot form, $\delta S_{bh} = 0$, and the
generalized second law reduces to $\delta S_{mat} \ge 0$, which is
satisfied by any physically reasonable equation of state.

Perturbations with wavelengths shorter than the Hubble length will
tend to collapse and form black holes via the Jeans instability.  Thus
consistency with the generalized second law suggests a holographic
bound may hold for regions smaller than the Hubble volume.

If $\omega$ is constant the particle horizon, $d_H$, and $1/H$ are
related to one another by a factor of order unity, and we recover the
Fischler-Susskind formulation.  However, if inflation has taken place
the particle horizon is much larger than $1/H$, which depends only
on the instantaneous expansion rate, and not on the integrated history
of the universe.
As we will see later, the Hubble volume is the relevant region to
consider in this case.

In principle, any initial value of $S/A$ is consistent with the
generalized second law.  However, if $S/A<1$ initially, this condition
is satisfied at all later times, provided $\omega$ is fixed and less
than unity \cite{FischlerET1998a}. With some additional assumptions,
we can also bound $S/A$ at early times.

As an example, consider the energy density and entropy density of
a relativistic gas at temperature $T$:
\begin{eqnarray}
\rho &=& \frac{\pi^2}{30} n_* T^4, \\
s &=& \frac{2 \pi^2}{45} n_* T^3,
\end{eqnarray}
where $n_*$ is the number of bosonic degrees of freedom plus $7/8$ times
the number of fermionic degrees of freedom. Using equation~(\ref{rw1})
to relate $\rho$ and $H$,  we find
\be
\frac{S}{A} \le \sqrt{n_*} T,
\ee
up to constant factors.  Since the density must be less than unity for
quantum gravitational corrections to be safely ignored, the maximum
temperature is proportional to $n_*^{-1/4}$, and the maximum value of
$S/A$ inside a Hubble volume is proportional to $n_*^{1/4}$.
Violating $S/A<1$ significantly at a sub-Planckian energy requires an
enormous value of $n_*$.  Thus, in the absence of fine tuning the
holographic bound originally proposed by Fischler and Susskind is
satisfied at all post-Planckian times.

\subsection{Closed Universe}

For an isotropic closed universe with fixed equation of state,
Fischler and Susskind found \cite{FischlerET1998a} that even if
$S/A<1$ initially on particle horizon sized regions, it could be
violated at later times.  This violation is possible even while
the universe is still in its expansion phase.

On the other hand, the generalized second law is expected to hold over
regions the size of the particle horizon in a closed universe. To be
definite, suppose this violation occurs while the universe is still in
its expansion phase. One certainly expects that a region with an
excessive entropy density could begin to collapse via the Jeans
instability and form a black hole. However, if the size of this region
is similar to that of the particle horizon, its collapse will
necessarily take at least a Hubble time. Consequently, a violation of
the holographic bound of Fischler and Susskind remains consistent with
the second law for cosmologically long time-scales.

Of course, in order to find $S/A>1$ well before the closed universe
reaches its final singularity, we must consider a volume that is a
substantial fraction of the overall universe. The collapse of this
region into a single black hole is not a small perturbation of the
background Friedmann universe.  Thus, the evolution equations for the
unperturbed universe are not expected to accurately describe the
entropy density of the collapsing region. 

\subsection{Open Universe}

The behavior of isotropic open (negative spatial curvature)
universes is similar to that of isotropic flat universes. If $S/A<1$ initially,
it remains so at later times \cite{FischlerET1998a}.  An argument that
$S/A<1$ remains valid at earlier times can likewise be made in a
similar way to the flat case.

\subsection{Inside a Black Hole}

Another interesting time dependent background is the region inside the
event horizon of a black hole.  The generalized second law will apply
to a spatial volume inside the event horizon if the volume is out of
thermal contact with other regions.  However it is straightforward to
argue that the entropy can exceed the surface area in such a volume
whose size is on the order of the horizon size. Consider a large ball
of gravitationally collapsing dust. The entropy of the ball is
approximately constant during the course of the collapse. However, the
size of the ball can contract to zero at the singularity, giving rise
to a violation of the holographic bound.

We see no reason why an observer inside such a region should not be
able to actually measure a violation of the holographic bound. A
direct measurement is difficult since the observer will typically hit
the singularity within a light-crossing time of the black hole
horizon.  However, if the observer has the additional information that
the entropy density is constant, s/he can infer a violation of
holography via local measurements.

\subsection{The Inflationary Universe}

We can view an inflationary universe as a Friedmann universe with a
time dependent equation of state.  During the reheating phase at the
end of inflation there is a sharp change in the equation of state, as
energy is transferred from the inflaton field to radiation (or
ultra-relativistic particles).  This raises the entropy of the
universe in a homogeneous way. After this sudden increase in the
entropy density it is possible to violate Fischler and Susskind's
bound when it is applied to regions the size of the particle
horizon. Of course, a sharp homogeneous increase in the entropy
density is permitted by the generalized second law.

The process of reheating is model dependent. To simplify the
discussion assume that reheating takes place instantaneously.  After
reheating, the equation of state need not change significantly. The
post-inflationary universe closely resembles a homogeneous and
isotropic universe which never inflated, with the exception that the
particle horizon of the universe after inflation is much larger than
that of a universe which did pass through an inflationary phase.  We
may therefore adopt the results for Friedmann universes with a
constant equation of state.

The post-inflationary universe is accurately approximated by an
isotropic Friedmann spacetime, so if $S/A<1$ when measured over a
Hubble volume at the end of inflation, this inequality will continue
to be satisfied at later times.  Moreover, immediately after reheating
the energy density is typically well below the Planck scale so $S/A
\ll 1$ in the absence of extreme fine-tuning, as discussed above.
This bound differs from that of Rama and Sarkar \cite{RamaET1998a},
and we obtain no specific constraints on inflationary models beyond
the usual assumption that the energy density is sub-Planckian during
and after inflation.

\section{Conclusions}

We have proposed that the holographic principle should be replaced by
the generalized second law of thermodynamics \cite{Bekenstein1973a} in
time dependent backgrounds. In static backgrounds, the
generalized second law reduces to the holographic principle of 't
Hooft and Susskind \cite{thooft, Susskind1995a}.  In cosmological
backgrounds corresponding to isotropic flat and open universes with a
fixed equation of state the second law implies the entropy bound of
Fischler and Susskind \cite{FischlerET1998a} over regions the size of the
particle horizon. However for closed universes, and inside black hole
event horizons, a useful holographic bound cannot be deduced from the
second law.  Finally, we proposed a modified version of the
holographic bound which applies to spatial regions of the
post-inflationary universe that are smaller than the Hubble volume.

\section*{Acknowledgments}

We thank Robert Brandenberger for useful comments.
This work was supported in part by DOE grant DE-FE0291ER40688-Task A.

\end{document}